\documentclass[amsmath,amssymb,superscriptaddress,twocolumn]{revtex4-1}

\usepackage{graphicx}
\usepackage{dcolumn}
\usepackage{bm}

\usepackage[colorlinks=true, citecolor=blue, urlcolor = blue, linkcolor= red, bookmarks=true]{hyperref}
\begin{document}

\title{ Nonsingular black hole chemistry in $4D$ Einstein-Gauss-Bonnet gravity }

\author{Arun Kumar}
\email{arunbidhan@gmail.com}
\affiliation{Centre for Theoretical Physics,
Jamia Millia Islamia, New Delhi 110025,
India}
\affiliation{Department of Mathematical Science, University of Zululand, Private Bag X1001, Kwa-Dlangezwa 3886, South Africa} 
\author{Sushant G. Ghosh}
\email{sghosh2@jmi.ac.in }
\affiliation{Centre for Theoretical Physics,
Jamia Millia Islamia, New Delhi 110025,
India} 
\affiliation{Astrophysics and Cosmology Research Unit,
School of Mathematics, Statistics and Computer Science,
University of KwaZulu-Natal, Private Bag X54001,
Durban 4000, South Africa}

\begin{abstract}
The EGB is an outcome of quadratic curvature corrections to the Einstein-Hilbert gravity action in the form of a Gauss-Bonnet (GB) term in $ D > 4$ dimensions, and EGB gravity is topologically invariant in $4D$. Several ways have been proposed for regularizing the $ D \to 4 $ limit of EGB for non-trivial gravitational dynamics in $ 4D $. Motivated by the importance of AdS/CFT,  we obtain an exact static spherically symmetric nonsingular black hole in $4D$ EGB gravity coupled to the nonlinear electrodynamics (NED)  in an AdS spacetime. We interpret the negative cosmological constant $\Lambda$ as the positive pressure, via $ P=-\Lambda/8\pi$,  of the system's thermodynamic properties of the nonsingular black hole with an AdS background.  We find that for  $P<P_c$, the black holes with $C_P>0$ are stable to thermal fluctuations and unstable otherwise.  We also analyzed the Gibbs free energy to find that the small globally unstable black holes undergo a phase transition to the large globally stable black holes. Further, we study the $P-V$ criticality of the system and then calculate the critical exponents to find that our system behaves like Van der Walls fluid.  
\end{abstract}

\maketitle

\section{Introduction} 
Black holes are perhaps one of the fascinating objects in nature, predicted theoretically by Einstein's general relativity (GR).  The presence of curvature singularity surrounded by the event horizon \cite{Hawking:1973uf} is believed to be the drawback of GR, which can be resolved through a more fundamental theory of gravitation, say, quantum gravity. As the quantum gravity is yet to be developed, the attempts to solve the singularity problem within the classical gravity itself, kicked off in 1968 by the idea of the regular black hole by Bardeen \cite{BardeenReg}, in which he proposed a model for black holes having horizons but no curvature singularity. However, the source of the Bardeen black hole solution, which is nonlinear electrodynamics (NED), was given by Ayon-Beato and Garcia \cite{AyonBeato:2000zs}. There has been an enormous advance in the analysis and application of regular black holes, and several exciting papers appeared uncovering properties\cite{regular,Hayward:2005gi,Ghosh:2014hea}.  
 
 The investigation of black hole's thermodynamics has communicated a deep and fundamental relationship among gravitation, thermodynamics, and quantum theory. It was possible by using classical and semiclassical analysis and has been given rise to most of our present physical understandings of the nature of quantum phenomena in the strong gravity regime \cite{Wald:1999vt,Israel:1967wq}. Furthermore, the advancement of the quantum field theory in curved surfaces leads to the correspondence of the surface gravity to its temperature \cite{Hawking:1974sw} and its event horizon area to its entropy \cite{Bekenstein:1973ur}. The pioneering work of discovering phase transition between the Schwarzschild AdS black holes and thermal AdS space by hawking and Page \cite{Hawking:1982dh} has attracted astrophysicists towards studying black holes thermodynamics in AdS spacetimes. Notably, in the last few years, the treatment of negative cosmological constant as the thermodynamical pressure $P$ with conjugate thermodynamical volume $V$ has modified the first law of black hole thermodynamics by including new term $V dP$. Now, the black hole mass is playing the role of enthalpy \cite{jmd} instead of internal energy. This inclusion of the new $V dP$ term in the first law of black hole thermodynamics has resolved the problem of inconsistency of black hole thermodynamics with Smarr relation. Over a long time, it is a well-established fact that asymptotic charges uniquely specify black holes because of several black hole theorems. Nevertheless, more recently, motivations from higher dimensions, string theory and holography have lead to considerations that violate some of the assumptions of these black hole theorems.  
 
 A quadratic correction to Einstein-Hilbert action described by the Gauss-Bonnet term leads to Einstein–Gauss-Bonnet (EGB) theory which  is also a string-generated gravity theory when approaching the low energy limit.  It is a natural and effective generalization of GR to higher dimensions, was discovered by Lanczos \cite{Lanczos:1938sf} and rediscovered by David Lovelock \cite{Lovelock:1971yv}. The EGB theory provides a broader setup to explore several conceptual issues of gravity, and the theory is also free of ghosts \cite{Boulware:1985wk}, since its equations of motion have no more than the second derivative of the metric. The first black hole solution in $5D$ EGB gravity was obtained by Boulware and Desser \cite{Boulware:1985wk}, since then a lot of papers in EGB gravity, including their formation and thermodynamics \cite{13} have been presented in the literature. It is well known that the contribution of GB term to the equations of motion is directly proportional to $D-4$; hence it does not contribute in $4D$ case. This issue was resolved by Glavan and Lin \cite{Glavan:2019inb} by rescaling the GB constant as $\alpha/(D-4)$ bypassing the conditions of Lovelock's theorem \cite{Lovelock:1971yv}.  A similar regularization procedure was initially proposed by Tomozawa \cite{Tomozawa:2011gp}  with finite one-loop quantum corrections to Einstein gravity,  and they also found the spherically symmetric black hole solution, which results in the repulsive nature of gravity at short distances. Later  Cognola {\it et al.} \cite{Cognola:2013fva} reformulated the approach of Tomozawa \cite{Tomozawa:2011gp}  that mimic quantum corrections due to a GB invariant within a classical Lagrangian approach. 

Nevertheless,  several researchers attracted  to analyse this $4D$ EGB theory, in particular,
the spherically symmetric black hole solution obtained by Glavan and Lin \cite{Glavan:2019inb}  was extended to include charge \cite{Fernandes:2020rpa}, a nonstatic Vaidya-like 
radiating black hole in Ref.~\cite{Ghosh:2020vpc,Ghosh:2020syx}, black holes  coupled with NED \cite{Kumar:2020uyz,Kumar:2020bqf,Kumar:2020xvu},  with  cloud of string background \cite{Singh:2020nwo}.  The  black hole stability and quasi-normal modes  are also widely discussed
\cite{Konoplya:2020bxa,Churilova:2020aca,Mishra:2020gce,Yang:2020czk,Zhang:2020sjh,Aragon:2020qdc},  and  the rotating counter of the black holes  obtained \cite{Wei:2020ght,Kumar:2020owy}. 
Investigation of relativistic star \cite{Doneva:2020ped},  derivation of regularised field equations \cite{Fernandes:2020nbq},  Morris-Thorne like wormholes \cite{Jusufi:2020yus}, thermodynamics \cite{HosseiniMansoori:2020yfj},  gravitational lensing  by a black hole \cite{Islam:2020xmy,Jin:2020emq,Heydari-Fard:2020sib}, and  
generalisation to  Lovelock gravity \cite{Konoplya:2020qqh} were also  explored.  

In this paper,  we  discuss how higher curvature
corrections alter black hole solutions and their qualitative features from our knowledge of black holes in GR.   Hayward \cite{Hayward:2005gi,Frolov:2016pav} proposed, Bardeen-like, regular spacetimes describe the formation of a black hole from an initial vacuum region with a  finite density and pressures, vanishing rapidly at large small and behaving as a cosmological constant at a small distance. The formation and evaporation process of a  Hayward \cite{Frolov:2016pav} regular black hole could be explored up to a minimum size $l$.  Interestingly, the causal structure of the nonsingular Hayward black hole makes a resemblance with that of the Reissner-Nordstr\"{o}m spacetime \cite{DeLorenzo:2014pta}. Subsequently, the Hayward black hole has applications in the broad context of physical phenomenon \cite{Toshmatov:2015wga,Lin:2013ofa,DeLorenzo:2014pta,Neves:2019ywx,Contreras:2018gpl,Frolov:2016pav}.
The generalization of these static black holes to the axially symmetric case, Kerr-like black hole, was also addressed \cite{Bambi:2013ufa, Amir:2015pja, Ghosh:2014hea}. 

 It is the purpose of this paper to investigate the Hayward-like solutions describing nonsingular
black holes in the  $4D$ EGB theory of gravity in AdS spacetimes i.e nonsingular--AdS EGB black holes. We shall analyse, in turn,  the critical behaviour of the thermodynamic quantities, $P-V$ criticality and demonstrate that there exists a phase transition and critical phenomena similar to the ones in a Van der Waals liquid-gas system. Thus, it is our goal to connect nonsingular
black holes in the  $4D$ EGB theory of gravity with the concept of black hole chemistry \cite{Kubiznak:2014zwa,Kumar:2020cve,Tzikas:2018cvs}. The black hole chemistry  - a new perspective on black hole thermodynamics, with its different interpretation of black hole mass and the cosmological constant $\Lambda$ as a pressure term, has led to a new understanding of Van der Waals fluids,  phase transitions, and triple points,  behaviour from a gravitational viewpoint \cite{Kubiznak:2014zwa}. 

The paper is organized as follows.  In Sec. \ref{HAdS}, we obtain exact nonsingular--AdS black hole solution to $4D$  EGB gravity, which allows us to discuss the properties of black holes. The thermodynamic properties of the solution derived and discussed  in 
Sec.~\ref{Thermo}, for nonsingular--AdS EGB black holes.  
In Sec. \ref{spv} The thermodynamical stability analysis and $P-V$ criticality of the black holes has been discussed, and the critical exponents have been calculated in Sec. \ref{critic}. The paper ends with concluding remarks 
in Sec. \ref{con}.

\section{ $4D$ EGB gravity and nonlinear electrodynamics}
\label{HAdS}
 Here, we are interested in the second-order Lovelock gravity known as Einstein-Gauss-Bonnet gravity; hence we start with the following action of EGB gravity with rescaled GB constant $\alpha \to \alpha/(D-4)$ \cite{Glavan:2019inb} minimally coupled to NED
\begin{equation}\label{action}
\mathcal{I}=\frac{1}{2}\int_{\mathcal{M}}d^{D}x\sqrt{-g}\left[  R +\frac{\alpha}{D-4}\mathcal{L}_{GB}-2\Lambda  \right]+ \mathcal{I_M},
\end{equation} where $R$ is curvature scalar, $\Lambda=-(D-1)(D-2)/2l^2$ is negative cosmological constant and $\mathcal{I_M}$ is the action of matter which is NED here, which can be written as \cite{Dehghani:2006ke,Kumar:2018vsm,Ghosh:2020tgy}
\begin{equation}
\mathcal{I_M}=\frac{1}{16\pi}\int d^{D}x \sqrt{-g} \mathcal{L}(F),
\end{equation} where $F=F_{ab}F^{ab}/4$, with $F_{ab}$ the field stress tensor, which in terms of corresponding gauge potential $A_a$ is defined as $F_{ab}=\partial_{a}A_{b}-\partial_{b}A_{a}$. $\mathcal{L}(F)$ is the Lagrangian density of the NED, which for the nonsingular black hole solutions we are interested in, is given as \cite{Kumar:2020xvu,Ghosh:2020tgy}
\begin{equation}
\mathcal{L}(F)=\frac{2(D-1)(D-2)\mu g^{D-1}}{(r^{D-1}+g^{D-1})^2}
\end{equation} where $\mu$ is a constant and $g$ is the magnetic monopole charge of NED field. Now, the variation of action (\ref{action}) leads to the following equations of motion \cite{Cai,Dehghani:2006ke,Kumar:2018vsm}
\begin{equation}
G_{ab}+\alpha H_{ab}=T_{ab}\equiv 2\left[\frac{\partial \mathcal{L}(F)}{\partial F} F_{ac}F^{c}_{b}-\frac{1}{4}g_{ab}\mathcal{L}(F) \right],\label{FieldEq}
\end{equation}where $T_{ab}$ is the energy momentum tensor for the NED field. Einstein tensor $G_{ab}$ and Lanczos tensor $H_{ab}$ \cite{Lanczos:1938sf,Kastor:2006vw} are
\begin{eqnarray}
G_{ab}&=&R_{ab}-\frac{1}{2}R g_{ab},\nonumber\\
H_{ab}&=&2\Bigr( R R_{ab}-2R_{ac} {R}{^c}_{b} -2 R_{acbd}{R}^{cd} - R_{acde}{R}^{cde}{_b}\Bigl)\nonumber\\&&~~~~~~~~~~~~~~~~~~~~~~~~~~~~~~~~-\frac{1}{2}\mathcal{L}_{GB} g_{ab}.
\end{eqnarray} Here, we want to obtain the static spherically symmetric black hole solutions of Eq. (\ref{action}), hence we consider the following metric \cite{Kumar:2020xvu,Ghosh:2020tgy}
\begin{equation}
ds^2=-f(r)dt^2+f(r)^{-1}dr^2+r^2d\Omega_{D-2}^2,\label{metric}
\end{equation} 
with $f(r)$ is the metric function to be determined and
\begin{equation}
d\Omega^2_{D-2}=d\theta_1^2+\sum_{i=2}^{D-2}\left[\prod_{j=2}^i \sin^2\theta_{j-1} \right]d\theta^2_i,
\end{equation} 
is the line element of a $(D-2)$-dimensional unit sphere \cite{Myers:1986un,XD}. We use metric (\ref{metric}) in Eq. (\ref{FieldEq}), and obtained the ({$r,r$}) equation of motion \cite{Ghosh:2014pga}, which in the limit $D\to4$, yields
\begin{eqnarray}
&&r^3f^{\prime}(r)+ \alpha\big(f(r)-1\big)\Big(f(r)-1-2rf^{\prime}(r)\Big)\nonumber\\&& + r^2\big(f(r)-1\big)-\frac{3}{l^2}r^4= -\frac{6\mu g^3 r^4}{(r^3+g^3)^{2}}.\label{rr1}
\end{eqnarray}  which integrates to 
\begin{equation}
f_{\pm}(r)=1+\frac{r^2}{2\alpha}\left(1\pm\sqrt{1+\frac{8M\alpha}{r^3+g^3}-\frac{4\alpha}{l^2}}\right),\label{fr}
\end{equation} where $M$, the constant of integration, is identified as the mass of the black hole. In the large asymptotic limit ($r\gg g$), the negative branch of the solution is the well-known Schwarzschild AdS black hole, whereas the positive branch corresponds to the Schwarzschild Ads black holes with negative mass, which is not physical. Further, in the GR limits ($\alpha \to 0$), the solution's negative and positive branch corresponds to Hayward Ads black holes, respectively, with positive and negative mass. Thus, hereafter we shall only take the negative branch of the solution (\ref{fr}) into consideration and shall call it nonsingular--AdS EGB black hole.

\begin{figure*}
\begin{tabular}{c c}
\includegraphics[width=0.5 \textwidth]{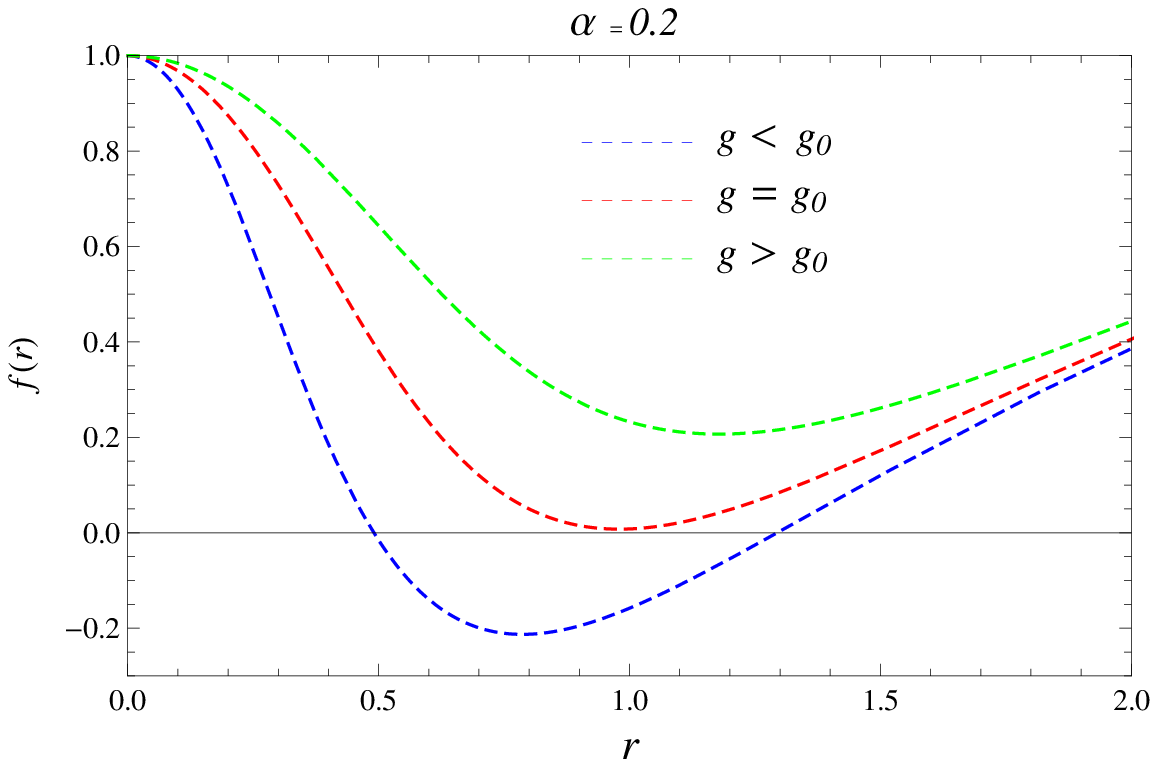}
\includegraphics[width=0.5 \textwidth]{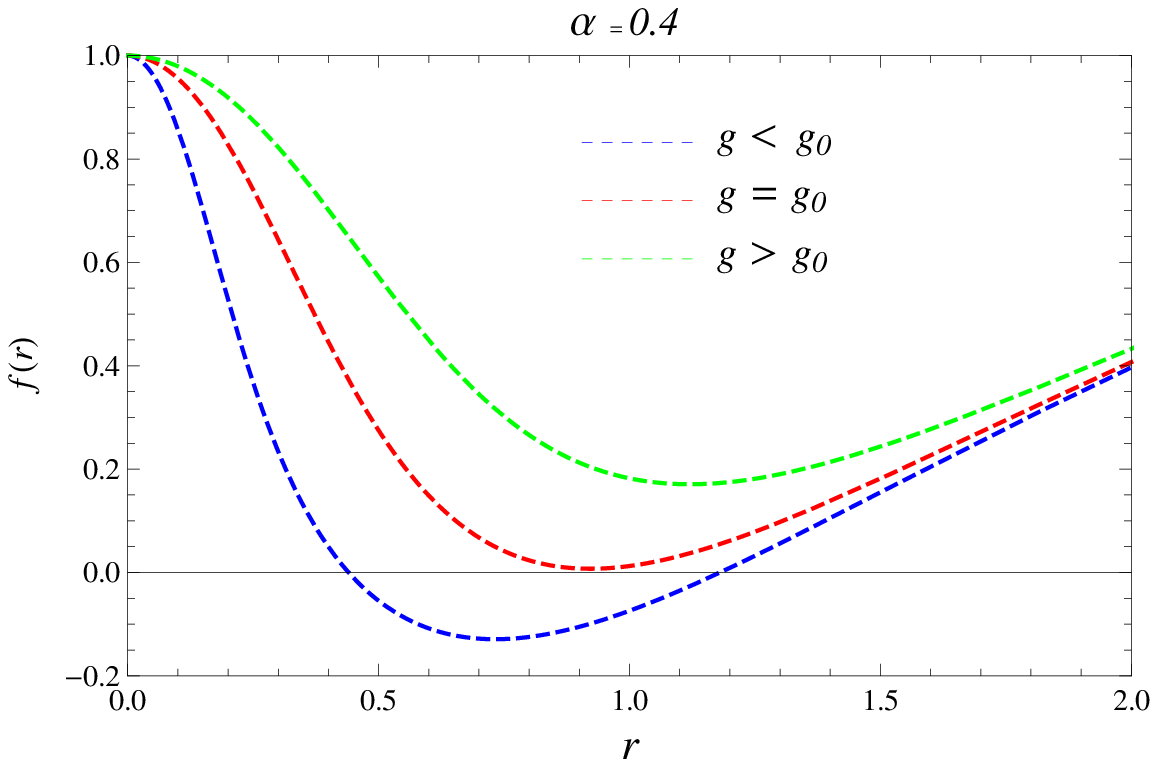}\\
\includegraphics[width=0.5 \textwidth]{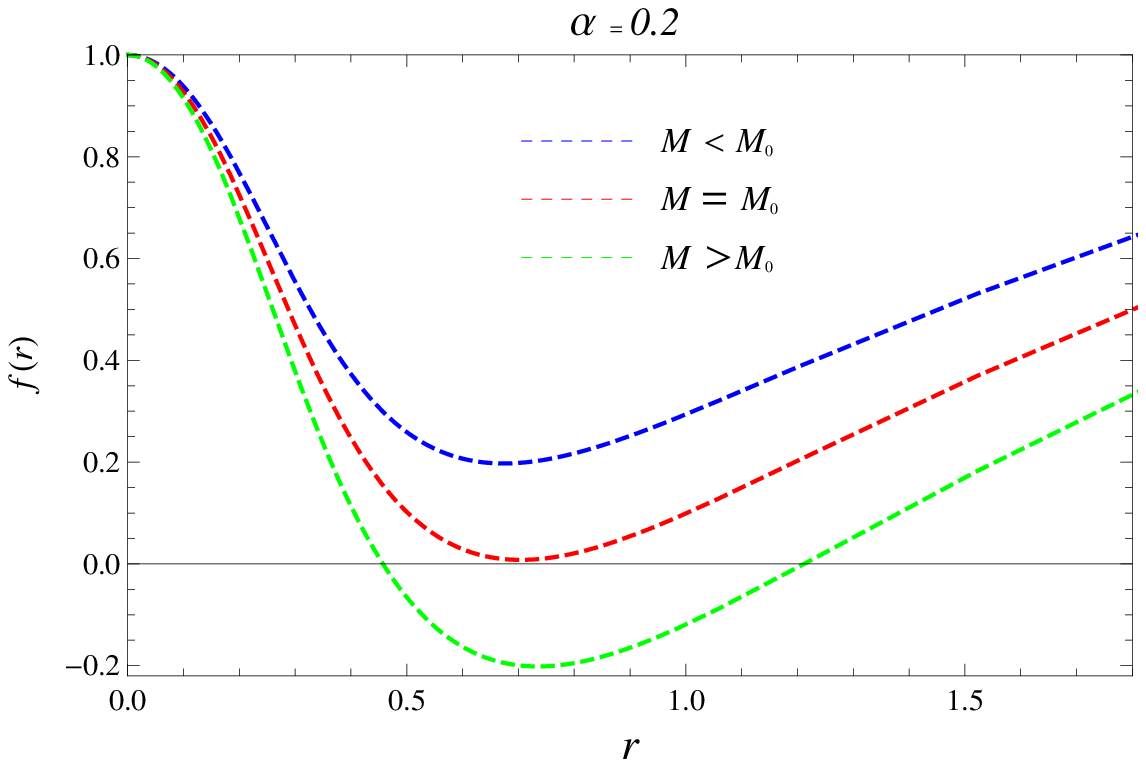}
\includegraphics[width=0.5 \textwidth]{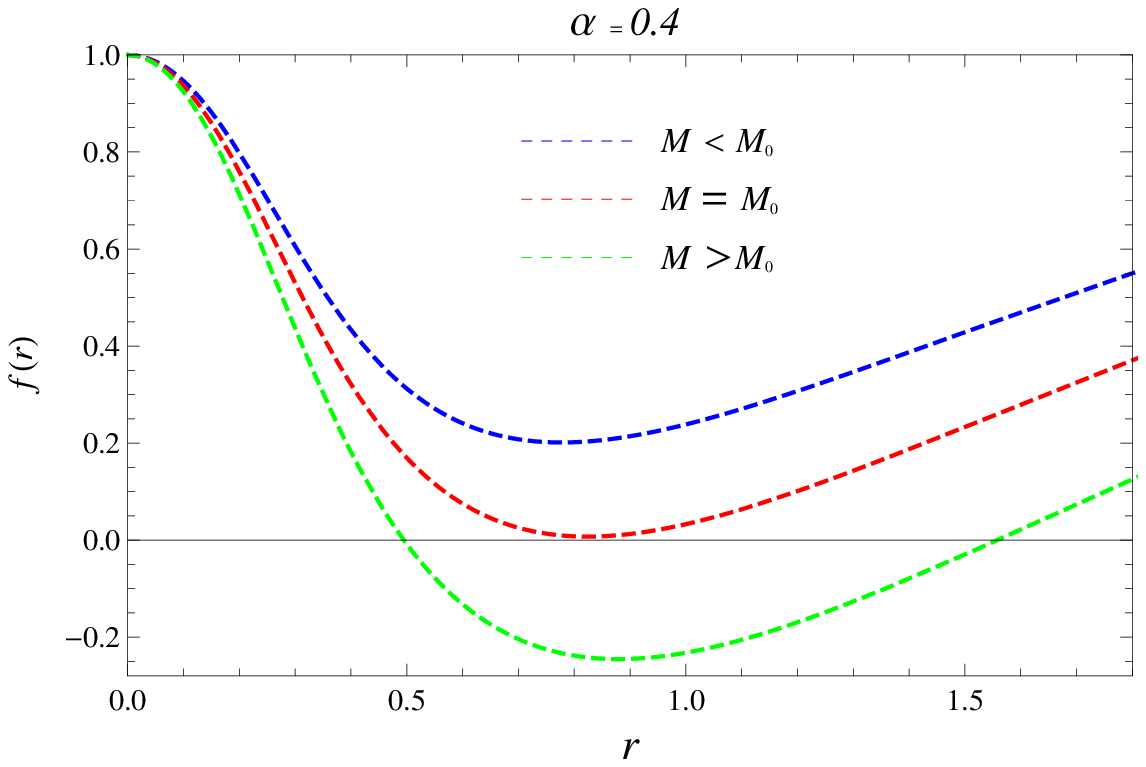}\\ 
\end{tabular}
\caption{ The metric function $f(r)$ for one-horizon and double-horizon nonsingular--AdS EGB black holes configurations.}
\label{fig:Bpotential}
\end{figure*}

The horizons of this nonsingular--AdS EGB black holes are given by  roots of the equation $g^{rr}(r_H)=0 \,\; \mbox{or} \,\;  f_{-}(r_H)=0$, which imply
\begin{equation}\label{horizon}
\left(r_H^3+g^3\right)\left(r_H^4+l^2r_H^2\right)-2Ml^2r_H^4=0.
\end{equation} 
 Obviously Eq.~(\ref{horizon}) may not be solved exactly 
and hence is depicted  in  Fig. \ref{fig:Bpotential}, by plotting  Eq. (\ref{horizon}) as a function of $r$ with intersection 
point on the $ r $-axis which gives two possible horizons.
Interestingly, there exist values of $M$ and $\alpha$ that has constrained the value of GB constant $g\geqslant g_0$ such that when $g>g_0$, there exists a black hole with two horizons $r_{\pm}$, where $r_-$ and $r_+$, respectively, are Cauchy and the event horizon. For $g=g_0$, both the horizons converge, and hence we get the extremal black hole with degenerate horizon $r_0=r_{\pm}$. Similarly, one can also get the maximum (minimum) allowed value of $\alpha$ ($M$) to obtain the black hole solution (cf. Fig. \ref{fig:Bpotential}).   Also, Fig. \ref{fig:Bpotential} suggests that there can be no black hole if the original mass is less than the minimal mass $M_0$. Furthermore, contrary to the usual case, there can be two horizons for large masses for $M>M_0$. As $M $ increases,  the inner horizon $r_- \to 0$, while the $r_+  \to 2 M$ the Schwarzschild radius similar to the noncommutative inspired black holes \cite{Nicolini:2005vd,Ghosh:2020cob}. 

The extremal black hole configuration is necessary because it corresponds to a thermodynamically stable remnant configuration and plays a fundamental role in the so-called gravity self-completeness paradigm \cite{Spallucci:2011rn}. The extremal black hole is characterised by \cite{Kumar:2020cve,Tzikas:2018cvs}
\begin{equation}
f(r)=0=\frac{\partial f(r)}{\partial r}|_{r=r_0} \,,
\end{equation}
which admits a solution 
\begin{equation}
M_0=\frac{\left(r_0^3+g^3\right)\left(\frac{r_0^4}{l^2}+r_0^2+\alpha\right)}{2r_0^4},
\end{equation} 
where $r_0$ is the horizon radius of the extremal black hole can be obtained from
\begin{equation}
3r_0^3\left[\frac{r_0^4}{l^2}+r_0^2+\alpha\right]-2\left(r_0^3+g^3\right)\left(r_0^2+2\alpha\right)=0.
\end{equation} 
In the absence of NED ($g=0$), the black hole is consistent with AdS--$4D$ EGB black hole \cite{Singh:2020mty}. 
\begin{center}
\begin{table}[h]
\begin{center}
\begin{tabular}{|l|l r  |l r  | }
\hline
\multicolumn{1}{|c|}{ }&\multicolumn{1}{c}{ }{$\alpha=0.1$}&\multicolumn{1}{c|}{ }&\multicolumn{1}{c}{ }{\,\,\,\,\,\,\,\,\,\,\,\,\,$\alpha=0.2$ }&\multicolumn{1}{c|}{}\\
\hline
\multicolumn{1}{|c|}{{$l$}}&\multicolumn{1}{c}{ $r_{0}$ } & \multicolumn{1}{c|}{ $M_{0}$} &\multicolumn{1}{c}{ $r_{0}$ } & \multicolumn{1}{c|}{ $M_{0}$} \\
\hline

\, $1$\,& \,\,\,\,\,\,\,0.52902\,\, &\,\,  0.62024\,\, &~~~~0.58137\,\,\,\,\,\,\,\,\,\,\,&\,\,0.74364~~~
\\
\
 \,$2$\,\, & \,\,\,\,\,\,\,0.59093\,\, &\,\,  0.53175\,\,&~~~~0.65821\,\,\,\,\,\,\,\,\,\,\,&\,\,0.63263~~~
\\
\
 \,$3$\,\, & \,\,\,\,\,\,\,0.61180\,\, &\,\,  0.51223\,\,&~~~~0.68627\,\,\,\,\,\,\,\,\,\,\,&\,\,0.60716~~~
\\
\
 \,$4$\,\, & \,\,\,\,\,\,\,0.50499\,\, &\,\,  0.69887\,\,&~~~~0.59753\,\,\,\,\,\,\,\,\,\,\,&\,\,0.04887~~~
\\
\
 \,$5$\,\, & \,\,\,\,\,\,\,0.62532\,\, &\,\,  0.50155\,\,&~~~~0.70541\,\,\,\,\,\,\,\,\,\,\,&\,\,0.59292~~~
\\
 \hline
\end{tabular}
\end{center}
\caption{The radius $r_0$ and mass $M_0$ for the extremal nonsingular--AdS EGB black holes ($g=0.4$).}
\label{tab0}
\end{table}
\end{center}
The numerical values of horizon radius $r_0$ and mass $M_0$ for nonsingular--AdS EGB black holes for various values of $l$ and $\alpha$ are tabulated in Table \ref{tab0} to find out that extremal black hole radius $r_0$ increases for increasing $l$ and $\alpha$. Whereas, the mass $M_0$ has the opposite behaviour. 

\section{Chemistry of nonsingular--AdS EGB black holes}
\label{Thermo}
\begin{figure*}
\begin{tabular}{c c c c}
\includegraphics[width=0.5 \textwidth]{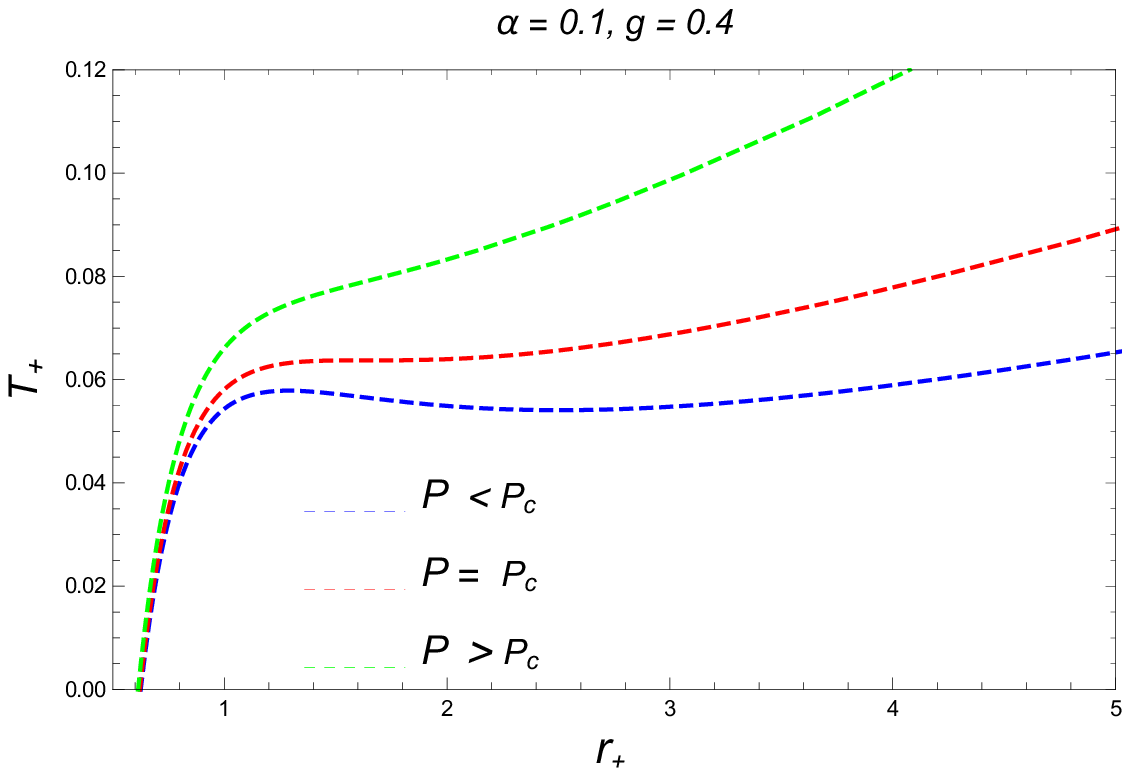}
\includegraphics[width=0.5 \textwidth]{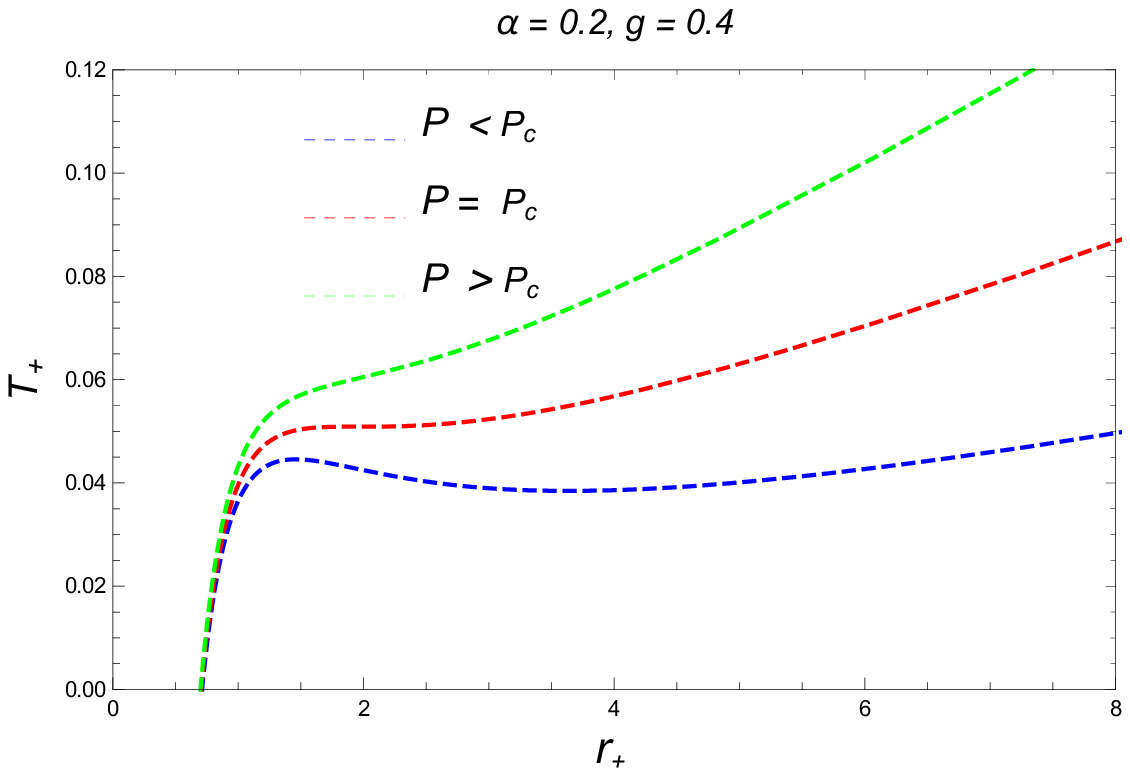}
\end{tabular}
\caption{ The Hawking temperature $T_+$ \textit{vs} horizon $r_+$ for nonsingular--AdS EGB black holes .}
\label{temp}
\end{figure*}

The black hole thermodynamics has witnessed significant attention, and  the role of cosmological constant as   thermodynamic pressure has been fleshed out and explored \cite{Kastor:2011qp,Kubiznak:2016qmn,Kubiznak:2014zwa,dr2,Kubiznak:2012wp}.   Also, the mass term in the first law of thermodynamics treated as enthalpy and not the internal energy of the black hole \cite{Dolan:2011xt,Dolan:2010ha}.   Tietelboim and Brown \cite{Teitelboim:1985dp,Brown:1987dd} were the first to purpose the interpretation of $\Lambda$ as a dynamical variable; however, the corresponding thermodynamical term was firstly incorporated in the first law of black hole thermodynamics in \cite{Creighton:1995au}. Then the generalised first law of black hole thermodynamics \cite{Kastor:2011qp,Kubiznak:2016qmn,Kubiznak:2014zwa,dr2,Kubiznak:2012wp} includes not only the charges  but also the impact of the non-zero energy coming from the cosmological constant in the volume inside the black hole
\begin{equation}\label{firstlw}
dM= TdS + VdP +\phi dg,
\end{equation} where $M (S,g,P)$ is black hole mass function with dynamical variables  the entropy $S$, the pressure $P$ and the magnetic charge $g$. While the temperature $T$, the thermodynamical volume $V$ and the chemical potential $\phi$, respectively, are the conjugate variables of $S$, $P$ and $g$ and which can be defined as \cite{Kastor:2011qp,Kubiznak:2016qmn,Kubiznak:2014zwa,dr2,Kubiznak:2012wp}
\begin{equation}
T=\left(\frac{\partial M}{\partial S}\right)_{g,P}, ~~~~\phi=\left(\frac{\partial M}{\partial g}\right)_{S,P}, ~~~~V=\left(\frac{\partial M}{\partial P}\right)_{g,S}.
\end{equation} 
The Smarr relation \cite{Smarr:1972kt}, due to cosmological term, generalized to \cite{Zhang:2018hms,Kubiznak:2014zwa}
\begin{equation}\label{smarr2}
M=2TS'-2PV+{\phi} g.
\end{equation}
It will be useful to understand the effect of magnetic charge on black hole thermodyanmics in canonical ensemble by considering magnetic monopole charge $g$ constant.  The black hole horizon satisfies $f(r_+)=0$, which gives black hole mass, in terms of $r_+$ as
\begin{equation} \label{B_mass}
M_+=H_+= \frac{r_+}{2} \left[\left(1+\frac{\alpha}{r_+^2}+\frac{8\pi P}{3}r_+^2\right)\left(1+\frac{g^3}{r_+^3}\right)\right].
\end{equation} The volume $V$ of the nonsingular--AdS EGB black holes can be obtained by using first law of black hole thermodynamics (\ref{firstlw})  
\begin{eqnarray}\label{V}
V=\frac{4}{3}\pi (r_+^3+g^3),
\end{eqnarray} which is independent of GB constant $\alpha$. By choosing appropriate limit $g=0$, one can get, $V=4 \pi r_+^3/3$, the volume of Schwarzschild black hole \cite{Tzikas:2018cvs}.
The Hawking temperature can also help us to understand the final stage  of the nonsingular--AdS EGB black hole evaporation. The Hawking temperature associated with the black hole is defined as $T=\kappa/2\pi$ \cite{Hawking:1974sw,Gibbons:1977mu,Gibbons:1976ue}, where $\kappa$ is the black hole surface gravity. Then, the temperature of nonsingular--AdS EGB black holes
\begin{equation} \label{B_temp0}
T_+= \frac{1}{4\pi r_+}\left[\frac{8 \pi P r_+^76+r_+^3(r_+^2-\alpha)-2g^3(r_+^2+2\alpha)}{(r_+^3+g^3)(r_+^2+2\alpha)}\right]. 
\end{equation} The isobars of nonsingular--AdS EGB black holes on the $T_+-r_+$ planes are depicted in the Fig. \ref{temp} for different values of the pressure, magnetic charge and GB coupling constant. For proper pressure say $P < P_c$, the isobaric curves have local maximum $T_{max}$ and local minimum $T_{min}$ at corresponding event horizon radii $r_{max}$ and $r_{min}$. Here, $r_{max}$ and $r_{min}$ are the smaller and the larger real positive roots of $\partial T_+/\partial r_+=0$. The numerical results of the local maximum temperature $T_{max}$ and the local minimum temperature $T_{min}$ with corresponding horizon radii $r_{max}$ and $r_{min}$ has been tabulated in Table \ref{tab1}, from which we can see that as one increases the value of $P$ or $\alpha$, the values of $T_{max}$, $T_{min}$ and $r_{max}$ increase whereas, the value of $r_{min}$ decreases.  When one takes the pressure $P=P_c$, the maximal and minimal temperature points coincide, i.e.,
\begin{equation}\label{inflex1}
\left(\frac{\partial T}{\partial r_+}\right)_{P_c}=\left(\frac{\partial^2 T}{\partial^2 r_+}\right)_{P_c}=0.
\end{equation} By using Eq. (\ref{B_temp0}) in Eq. (\ref{inflex1}), we get the following expression for critical pressure
\begin{widetext}
\begin{equation}
P_c=\frac{r_c^6\left[r_c^4-\alpha \left(5r_c^2+2\alpha\right)\right]-2g^3\left[5r_c^7+2\alpha r_c^3\left(10r_c^2+7\alpha\right)+g^3\left(r_c^2+2\alpha\right)^2\right]}{4\pi r_c^7\left[r_c^3\left(r_c^2+6\alpha\right)+4g^3\left(r_c^2+3\alpha\right)\right]},
\end{equation}
\end{widetext}where $r_c$ is the critical horizon radius, which is the larger positive root of Eq. (\ref{inflex1}). It is very complex to solve Eq. (\ref{inflex1}) analytically and hence we solve Eq. (\ref{inflex1})  in some limiting cases to get
$r_c(\alpha \to 0)=g\left[2(7+3\sqrt{6})\right]^{1/3}$ and $r_c(g \to 0)= \sqrt{\alpha(6+4\sqrt{3})}.$
The critical pressure $P_c$ in various limits
\begin{eqnarray}\label{pc1}
P_c(\alpha \to 0)&=&\frac{3^{5/3}}{16\pi (2g)^{2/3}\left(16911+6904\sqrt{6}\right)^{1/3}},\nonumber\\P_c(g \to 0)&=& \frac{15-8\sqrt{3}}{288\pi \alpha}.
\end{eqnarray}
  For $P>P_c$, one can clearly see that the black hole temperature is increasing monotonically.
\begin{widetext}
\begin{center}
\begin{table}[h]
\begin{center}
\begin{tabular}{|l|l r r l |l r r l | }
\hline
\multicolumn{1}{|c|}{ }&\multicolumn{1}{c}{ }&\multicolumn{1}{c}{ }&\multicolumn{1}{c}{ }{$\alpha=0.1$}&\multicolumn{1}{c|}{ }&\multicolumn{1}{c}{ }&\multicolumn{1}{c}{ }{\,\,\,\,\,\,\,\,\,\,\,\,\,$\alpha=0.2$ }&\multicolumn{1}{c}{}&\multicolumn{1}{c|}{}\\
\hline
\multicolumn{1}{|c|}{{$P$}}&\multicolumn{1}{c}{ $r_{max}$ } & \multicolumn{1}{c}{ $T_{max}$}&\multicolumn{1}{c}{{$r_{min}$}}&\multicolumn{1}{c|}{$T_{min}$}  &\multicolumn{1}{c}{ $r_{max}$ } & \multicolumn{1}{c}{ $T_{max}$}&\multicolumn{1}{c}{{$r_{min}$}}&\multicolumn{1}{c|}{~$T_{min}$} \\
\hline

\,\,\,\, $0.001$\,& \,\,\,\,\,\,\,1.15178\,\, &\,\,  0.04964\,\, &~~~~~~6.20666\,\,\,\,\,\,\,\,\,\,\,&\,\,0.02505& \,\,\,\,\,\,\,1.37214\,\, &\,\,  0.04112\,\, &~~~~~~6.11270\,\,\,\,\,\,\,\,\,\,\,&\,\,0.02489
\\
\
\,\, $0.002$\,\, & \,\,\,\,\,\,\,1.17702\,\, &\,\,  0.05159\,\,&4.30500\,\,\,\,\,\,\,\,\,\,\,&\,\,0.03516& \,\,\,\,\,\,\,1.42617\,\, &\,\,  0.04339\,\, &4.15804\,\,\,\,\,\,\,\,\,\,\,&\,\,0.03468
\\
\
\,\, $0.003$\,\, & \,\,\,\,\,\,\,1.20651\,\, &\,\,  0.05360\,\,&3.43606\,\,\,\,\,\,\,\,\,\,\,&\,\,0.04271& \,\,\,\,\,\,\,1.49897\,\, &\,\,  0.04580\, &3.22881\,\,\,\,\,\,\,\,\,\,\,&\,\,0.04176
\\
\
\,\, $0.004$\,\, & \,\,\,\,\,\,\,1.24209\,\, &\,\,  0.05569\,\,&2.89562\,\,\,\,\,\,\,\,\,\,\,&\,\,0.04887& \,\,\,\,\,\,\,1.61384\,\, &\,\,  0.04841\,\, &2.59564\,\,\,\,\,\,\,\,\,\,\,&\,\,0.04728      
\\
 \hline
\end{tabular}
\end{center}
\caption{The local maximum temperature $T_{max}$ and  the local minimum temperature $T_{min}$ corresponding to the horizon radius $r_{max}$ and $r_{min}$ for nonsingular--AdS EGB black holes.}
\label{tab1}
\end{table}
\end{center}
\end{widetext}
We plug in the values of black hole mass and temperature, respectively, from Eqs. \eqref{B_mass} and \eqref{B_temp0} in Eq. \eqref{firstlw} and integrate to get the entropy of the nonsingular--AdS EGB black holes
\begin{eqnarray} \label{B_s}
S_+&=&\frac{A}{4}
+2\pi \alpha \log\left(\frac{A}{A_0}\right)-\frac{2\pi g^3}{3r_+^3}\left(3r_+^2+2\alpha\right).\end{eqnarray} Here, $A=4\pi r_+^2$ is the black hole event horizon area, and $A_0$ is a constant with the area units. This equation generalizes the Hawking-Bekenstein area formula \cite{Bekenstein:1973mi} by a supplementary logarithmic term  and third in the above expression is due to magnetic charge $g$.
Notice that, in limit $\alpha \to 0$, we obtain the entropy  of the Hayward black hole  \cite{Kumar:2020xvu}. Here, it is interesting to note that entropy of nonsingular--AdS EGB black holes does not depend explicitly on the pressure $P$ of the system, but the horizon radius $r_+$ of the black hole depends on pressure $P$; hence the pressure of the black hole affects its entropy. 
 
\section{Stability and $P-V$ Crticality}
\label{spv}
\begin{figure*} 
\begin{tabular}{c c c c}
\includegraphics[width=0.5 \textwidth]{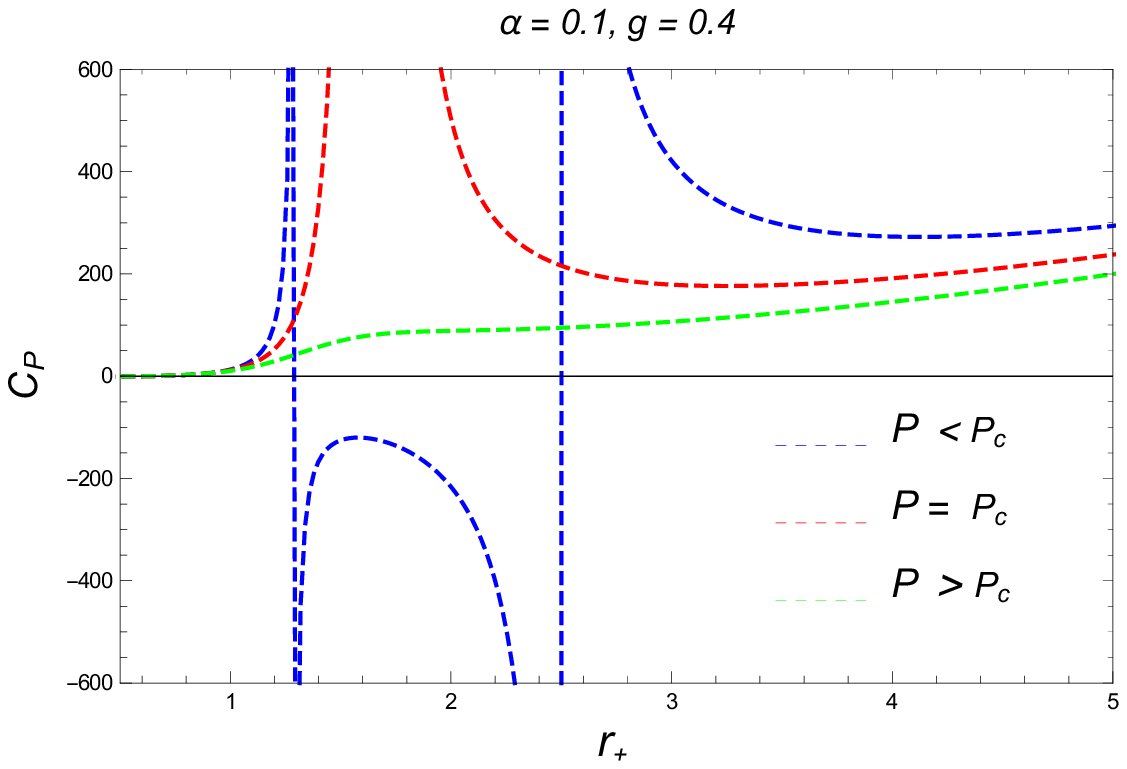}
\includegraphics[width=0.5 \textwidth]{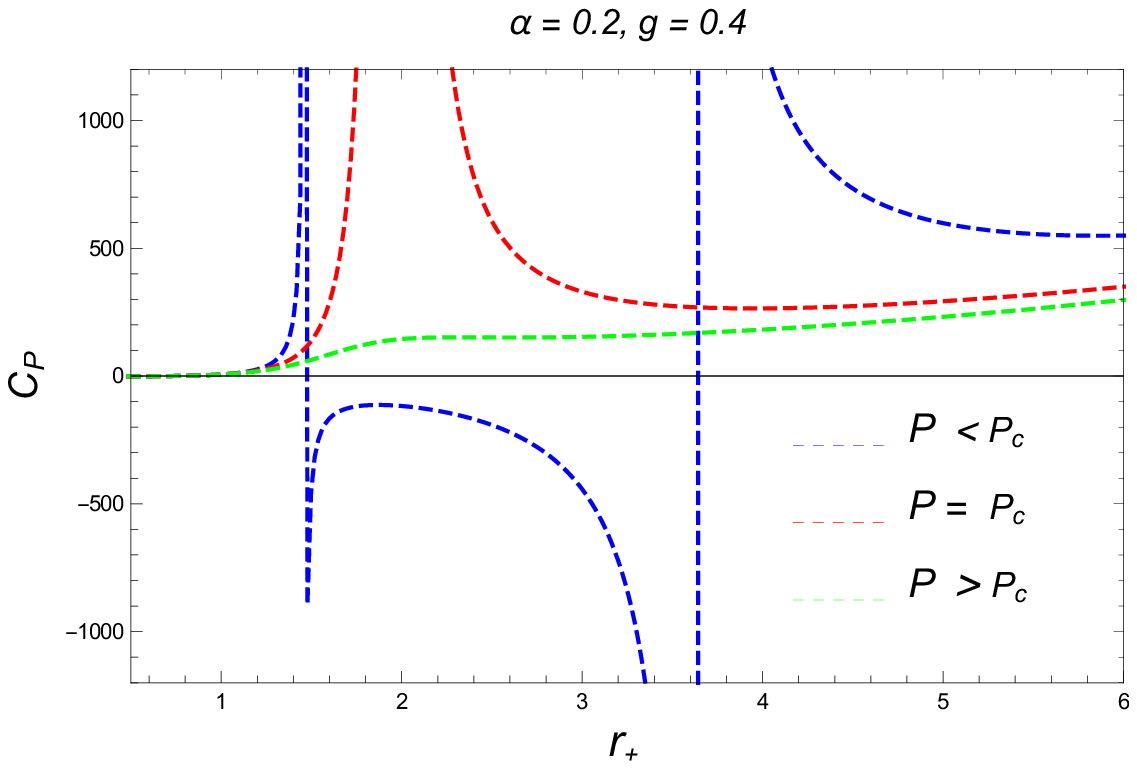}
\end{tabular}\caption{The heat capacity ($C_P$) \textit{vs} horizon $r_+$ for nonsingular--AdS EGB black holes.}
\label{heat}
\end{figure*}

 The black hole with positive heat capacity ($C_+>0$) is at least locally thermodynamical stable, whereas the heat capacity's negativity ($C_+<0$) signifies the thermodynamical instability of the black hole to the thermal fluctuations \cite{Sahabandu:2005ma,Cai:2003kt}. It turns out that at some stage, a black hole, due to thermal fluctuations, absorbs more radiation than it emits leading to positive heat capacity. Whereas when the black hole emits more radiation than it absorbs, the heat capacity becomes negative. By using $C_P = T \left({\partial S}/{\partial T} \right)_{P}$ \cite{Kumar:2020cve,Tzikas:2018cvs}, we get the heat capacity
 \begin{widetext}
\begin{eqnarray}
C_P=-2\pi r_+^2\left[\frac{\left(1+\frac{g^3}{r_+^3}\right)\left(r_+^2+\alpha\right)^2\left[8\pi P r_+^8+r_+^4\left(r_+^2-\alpha\right)-2g^3\left(r_+^2+2\alpha\right)\right]}{r_+^6\left[r_+^4-4\pi P r_+^4\left(r_+^2+6\alpha\right)-\alpha\left(5r_+^2+2\alpha\right)\right]-Ag^3-Bg^6}\right],
\end{eqnarray}
\end{widetext}with
$A=2r_+^3\left[16\pi P r_+^4\left(r_+^2+3\alpha\right)+5r_+^4+2r_+^2\left(10r_+^2+7\alpha\right)\right]$ and $B=2\left(r_+^2+2\alpha\right)^2$. The isobaric curves for different values of pressure $P$ and GB constant $\alpha$ are depicted in Fig. \ref{heat}. One can notice that when one takes pressure of the black hole less than $P_c$, the heat capacity suffers from discontinuities at $r_{max}$ and $r_{min}$ which divide curve in three different regions such that the $C_P>0$ for $r_+>r_{min}$ and $r_+<r_{max}$ and $C_P<0$ for $r_{max}<r_+<r_{min}$. Thus, the small and large black holes, respectively, with $r_+>r_{min}$ and $r_+<r_{max}$ are thermodynamically stable, whereas the intermediate black holes are unstable. The discontinuity of heat capacity at $r_{max}$ and $r_{min}$, confirms the existence of phase transition \cite{jd1}. It can be seen easily that the small (intermediate) black holes undergoes phase transition to intermediate (large) black holes at $r_{max}$ ($r_{min}$). When we take $P\geqslant P_c$, we find that the heat capacity is always positive signifying the local thermodynamical stability of the black holes.

\begin{figure*}
\begin{tabular}{c c c c}
\includegraphics[width=0.5 \textwidth]{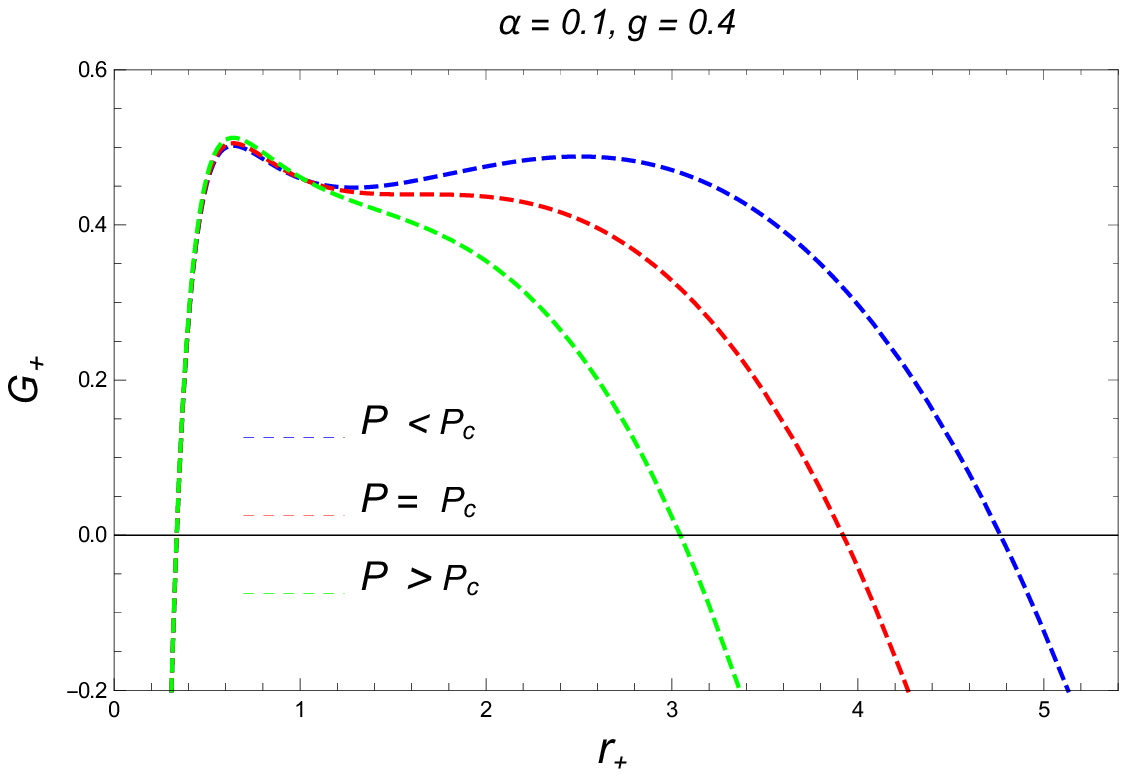}
\includegraphics[width=0.5 \textwidth]{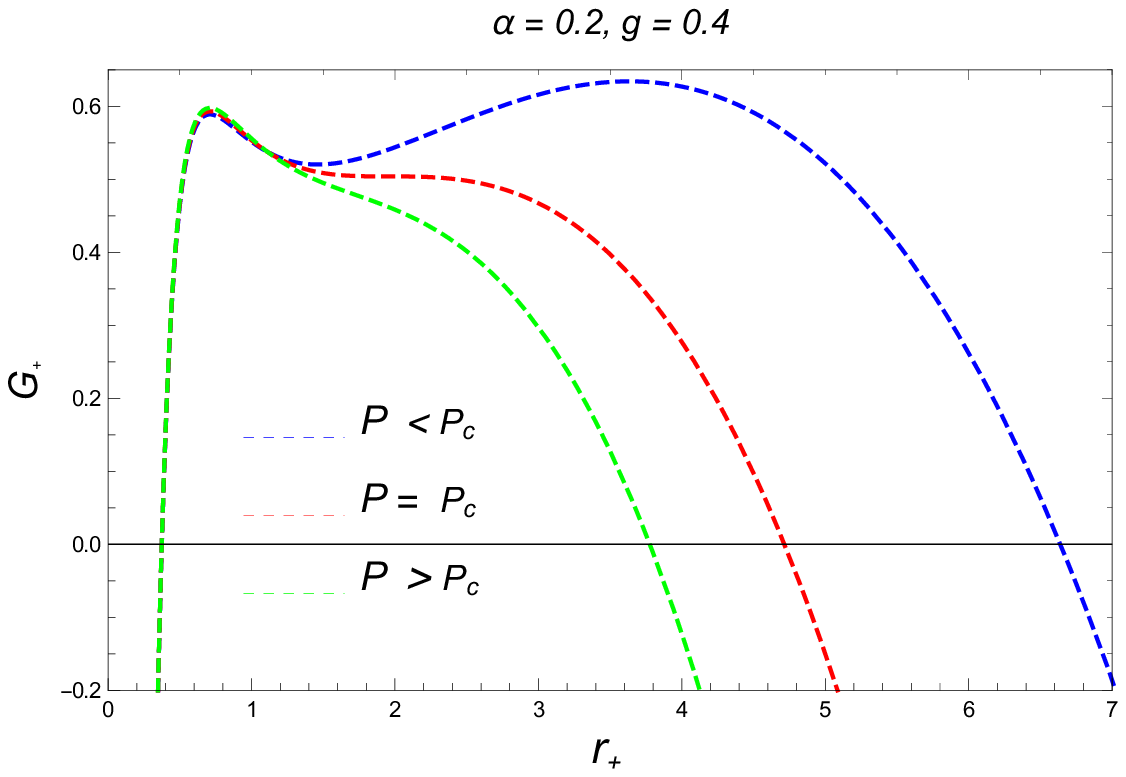}
\end{tabular}
\caption{ Gibb's free energy ($G_+$) \textit{vs} horizon $r_+$ for nonsingular--AdS EGB black holes.}
\label{gplot}
\end{figure*}

\begin{figure*} 
\begin{tabular}{c c c c}
\includegraphics[width=0.5 \textwidth]{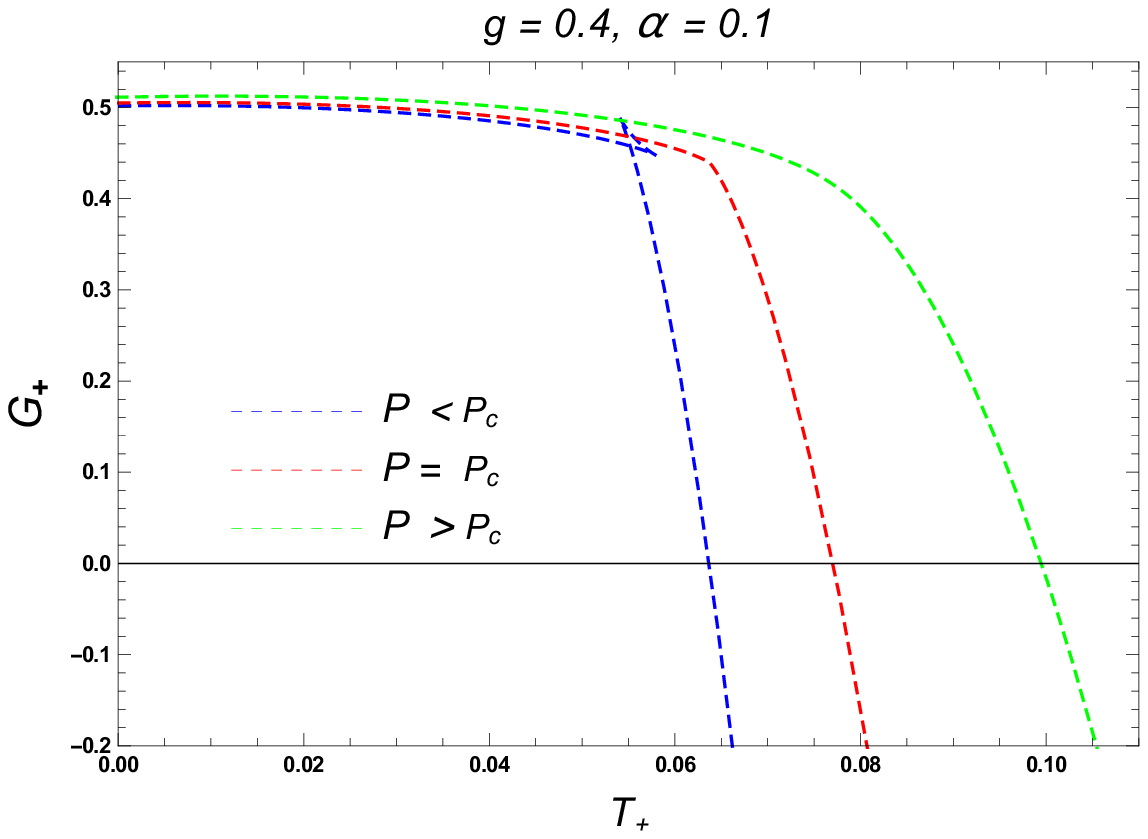}
\includegraphics[width=0.5 \textwidth]{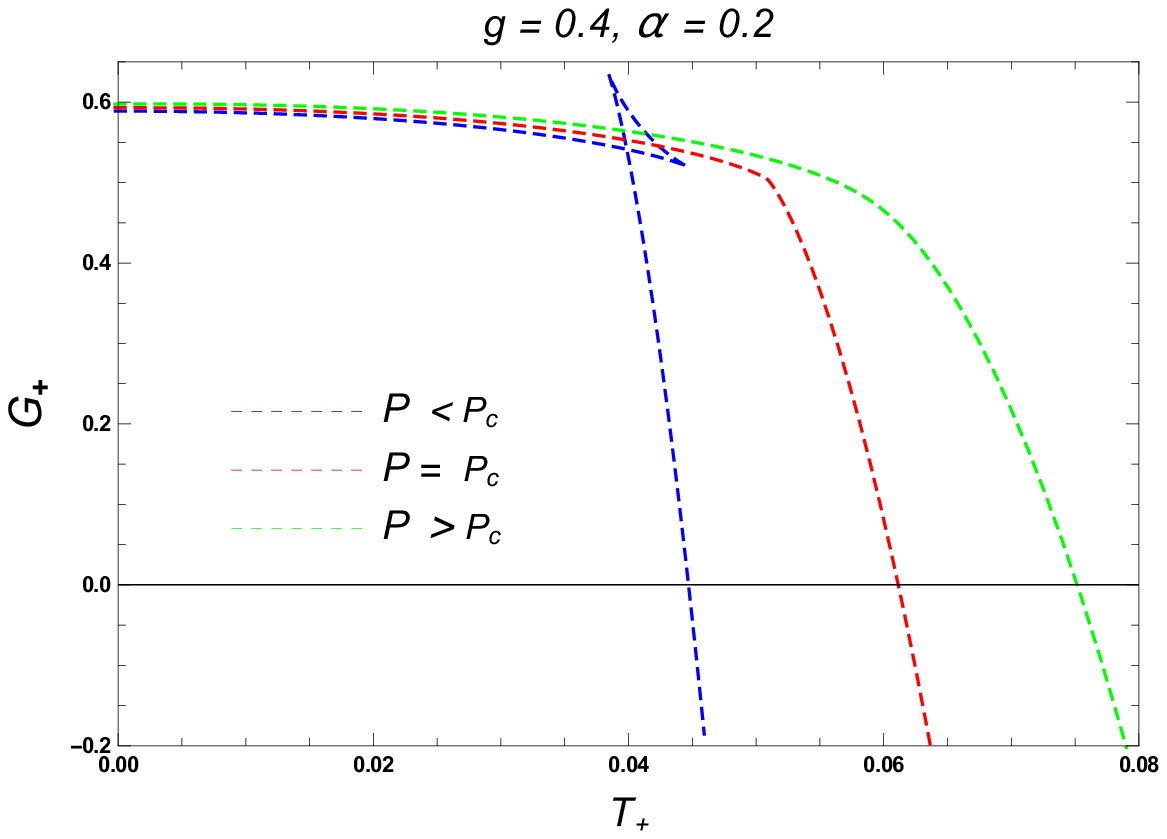}
\end{tabular}
\caption{ Gibb's free energy ($G_+$) \textit{vs} temperature $T_+$ for nonsingular--AdS EGB black holes.}
\label{gplot1}
\end{figure*}

The global stability of a black hole is governed by the sign of its free energy. The black holes with negative free energy ($G_+<0$) are globally stable. On the other hand, those with positive free energy are globally unstable. It is well known that in extended phase space, the mass of a black hole plays the role of enthalpy; hence the free energy of black, which is indeed Gibbs free energy $G_+$ can be defined as $G_+=M_+-T_+S_+$ \cite{Kumar:2020cve,Tzikas:2018cvs}.
The behaviour of Gibbs free energy $G_+$ vs. horizon radius $r_+$ and temperature $T_+$, for changing pressure $P$ and GB constant $\alpha$ has been plotted in Figs. \ref{gplot} and \ref{gplot1}. It is evident that the black holes with small and large horizon radii are globally stable with $G_+ <0$ (cf. Fig. \ref{gplot}). When one sees Fig. \ref{gplot1}, one notices that when $P<P_c$, the black hole undergoes the first-order phase transition between thermodynamically stable phases (swallowtail structure). However, for $P\geqslant P_c$, the first-order phase transition does not happen (no swallowtail structure). It can also be seen that the Gibbs free energy flips its sign ($+$ to $-$) at some particular value of temperature, say $T_{HP}$ with corresponding horizon radius $r_{HP}$, where it vanishes confirming the Hawking-Page phase transition between the black hole and thermal radiation \cite{Hawking:1982dh}. From the numerical results of temperature $T_{HP}$ and horizon radius $r_{HP}$ at Hawking-Page phase transition tabulated in Table \ref{tab2}, we conclude that the values of $T_{HP}$ increases while $r_{HP}$ decreases as we increase the value of pressure. But, for increasing $\alpha$ values of both $T_{HP}$ and $r_{HP}$ increase.
\begin{center}
\begin{table}[h]
\begin{center}
\begin{tabular}{|l|l r|r l |r l }

\hline
\multicolumn{1}{|c|}{ }&\multicolumn{1}{c}{ }{\,\,\,\,\,\,$\alpha=0.1$}&\multicolumn{1}{c|}{ }&\multicolumn{1}{c}{ }{\,\,\,\,\,\,\,\,\,\,\,\,\,$\alpha=0.2$ }&\multicolumn{1}{c|}{}\\
\hline
\multicolumn{1}{|c|}{{$P$}}&\multicolumn{1}{c}{ $r_{HP}$ } & \multicolumn{1}{c|}{ $T_{HP}$}&\multicolumn{1}{c}{{$r_{HP}$}}&\multicolumn{1}{c|}{$T_{HP}$}   \\
\hline

\,\,\,\, $0.003$\,& \,\,\,\,\,\,\,6.17804\,\, &\,\,  0.04963\,\, &6.03700\,\,\,\,\,\,\,\,\,\,\,&\,\,0.04877
\\
\
\,\, $0.005$\,\, & \,\,\,\,\,\,\,4.76783\,\, &\,\,  0.06367\,\,&4.63392\,\,\,\,\,\,\,\,\,\,\,&\,\,0.06213\,\,
\\
\
\,\, $0.007$\,\, & \,\,\,\,\,\,\,4.02894\,\, &\,\,  0.07499\,\,&3.90810\,\,\,\,\,\,\,\,\,\,\,&\,\,0.07277\,\,
\\
\
\,\, $0.009$\,\, & \,\,\,\,\,\,\,3.56126\,\, &\,\,  0.08474\,\,&3.45495\,\,\,\,\,\,\,\,\,\,\,&\,\,0.08188\,\,
\\
\
\,\, $0.011$\,\, & \,\,\,\,\,\,\,3.23419\,\, &\,\,  0.09345\,\,&3.14214\,\,\,\,\,\,\,\,\,\,\,&\,\,0.08999\,\,
\\
 \hline
\end{tabular}
\end{center}
\caption{The Hawking temperature $T_{HP}$ and horizon radius $r_{HP}$ at Hawking-Page phase transition point for nonsingular--AdS EGB black holes.}
\label{tab2}
\end{table}
\end{center}

To discuss the $P-V$ criticality of the nonsingular--AdS EGB black holes firstly we get the following equation of state $P=P(V,T)$ by using Eq. (\ref{B_temp0})
\begin{equation}\label{P}
P=\frac{(r_+^2+g^2)(r_+^2+2\alpha)T}{2r_+^5}+\frac{2g^2(r_+^2+2\alpha)-r_+^2(r_+^2-\alpha)}{4\pi r_+^6},
\end{equation} where $r_+$ is obviously a function of the volume of the system via $r_+=[(3V-2\pi g^3)/2\pi]^{1/3}$. We can find the expression for critical temperature $T_c$ which is the temperature of the system at the inflection point of the isotherm of Eq. (\ref{P}), through the following equations \cite{Kubiznak:2014zwa,Kubiznak:2012wp,Kubiznak:2016qmn}
\begin{equation} \label{crtical_eq}
\left(\frac{\partial P}{\partial r}\right)_{T} = 0 = \left(\frac{\partial ^2 P}{\partial r^2}\right)_{T} \,.
\end{equation} in terms of critical horizon radius $r_c$ as
\begin{equation}
T_c=\frac{r_c^3\left(r_c^2-2\alpha\right)-g^3\left(5r_c^2+14\alpha\right)}{2\pi r_c\left[r_c^3\left(r_c^2+6\alpha\right)+4g^3\left(r_c^2+3\alpha\right)\right]}
\end{equation}
We take limits $\alpha \to 0$ and $g\to0$, to get
\begin{eqnarray}\label{tc1}
T_c=\frac{5^{2/3}}{4\pi g\left(118+48\sqrt{6}\right)^{1/3}},\;T_c= \frac{1}{6\pi}\sqrt{\frac{-3+2\sqrt{3}}{2 \alpha}}.
\end{eqnarray}
\begin{figure*} 
\begin{tabular}{c c c c}
\includegraphics[width=0.5 \textwidth]{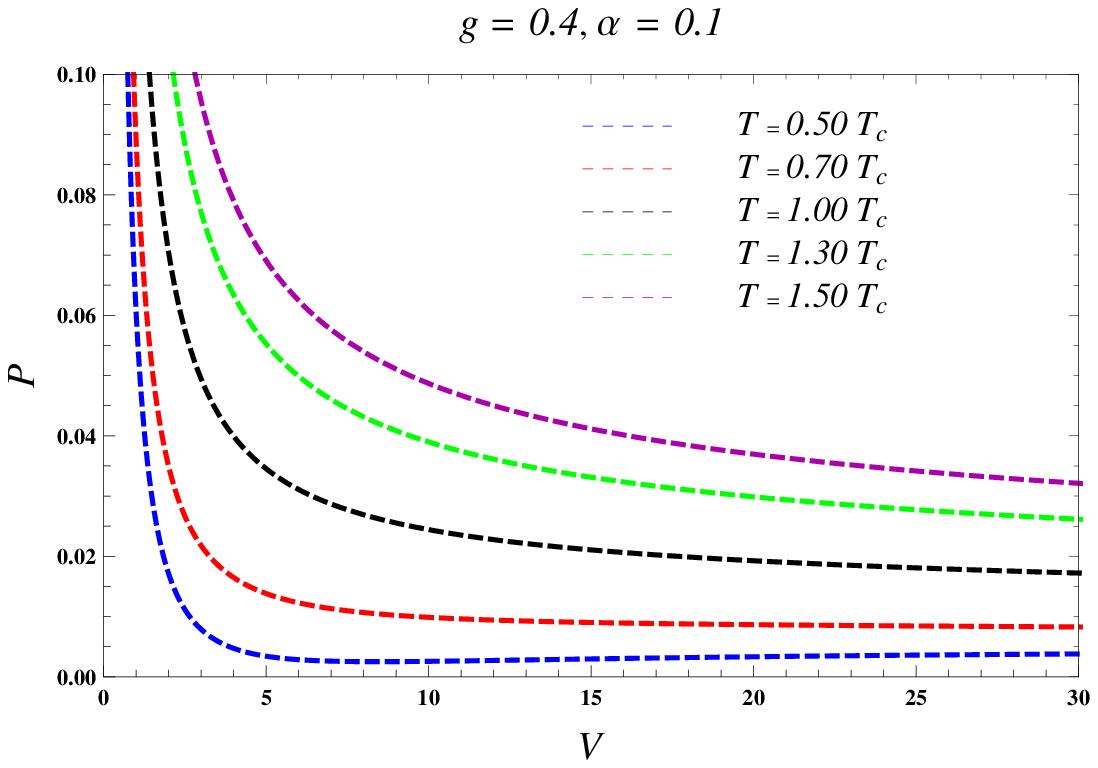}
\includegraphics[width=0.5 \textwidth]{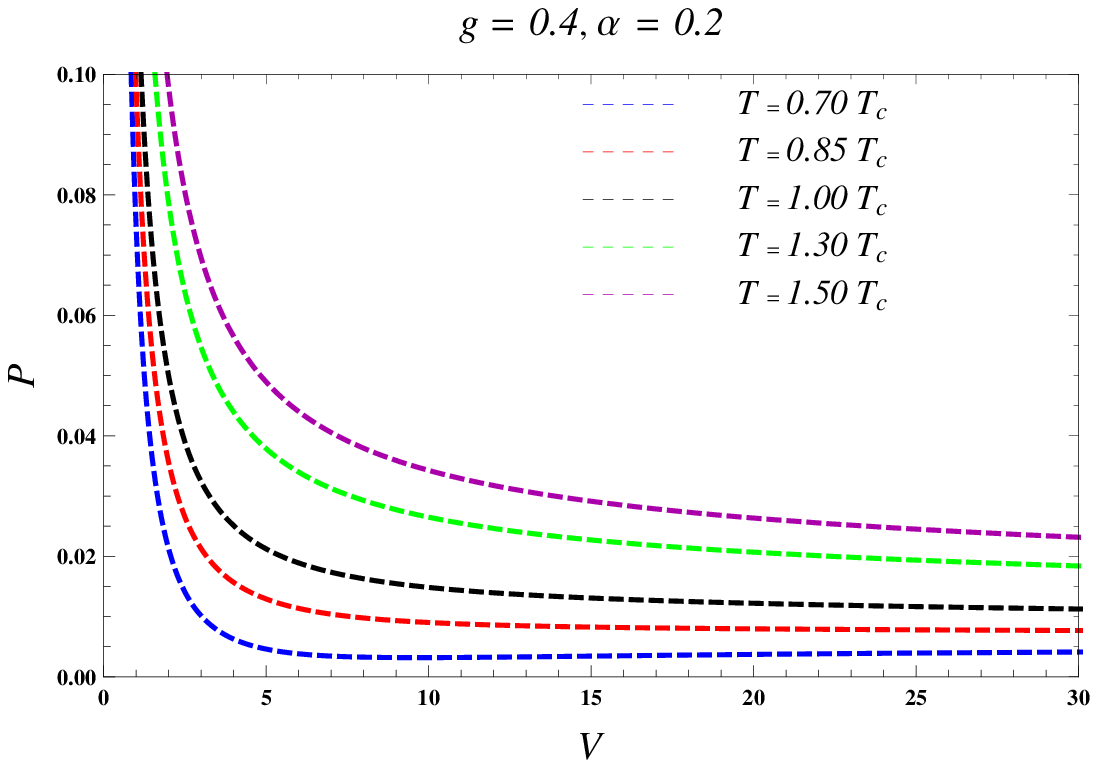}
\end{tabular}
\caption{ The pressure $P$ \textit{vs} the volume $V$ for nonsingular--AdS EGB black holes.}
\label{pvplot}
\end{figure*}
We plot the isotherms of Eq. (\ref{P}) in Fig. \ref{pvplot} to find out that when we take $T_+<T_c$ the isotherms are consisting of three different branches of black holes. The branch with small (large) pressure corresponds to small (large) black holes. There is an oscillating branch existing in between small and large black holes corresponding to intermediate thermodynamically unstable black holes, and hence the small black holes go over to the large black holes through Van der Wall like first-order phase transition. When we take $T>T_c$ there is no oscillating region as the pressure is constant with changing volume, and hence no more first-order phase transition happens. Now by using the expression for critical temperature $T_c$, pressure $P_c$, and the volume $V_c$, we can obtain the universal constant for nonsingular-AdS EGB black holes by using $\varepsilon = {(P_c V_c^{1/3})}/{T_c}$, whose analytical expression is omitted here. In the limiting cases $\alpha\to0$ and $g\to0$, the universal constant becomes $\epsilon\approx0.32$ and $\epsilon\approx0.29$, respectively. The numerical values of critical radius $r_c$, critical pressure $P_c$, critical temperature $T_c$ and the universal constant $\epsilon$ has been shown in Table \ref{tab3}, from which it can be concluded that value of $r_c$ increases but that of $P_c$ and $T_c$ decreases for increasing value of $g$ and $\alpha$. The value of universal constant $\epsilon$ increases and decreases, respectively, as the value of $g$ and $\alpha$ increases. It can be seen easily that the universal constant is less than that of Van der Walls fluid.
\begin{widetext}
\begin{center}
\begin{table}[h]
\begin{center}
\begin{tabular}{|l|l r r l |l r r l | }
\hline
\multicolumn{1}{|c|}{ }&\multicolumn{1}{c}{ }&\multicolumn{1}{c}{ }&\multicolumn{1}{c}{ }{$\alpha=0.1$}&\multicolumn{1}{c|}{ }&\multicolumn{1}{c}{ }&\multicolumn{1}{c}{ }{\,\,\,\,\,\,\,\,\,\,\,\,\,$\alpha=0.2$ }&\multicolumn{1}{c}{}&\multicolumn{1}{c|}{}\\
\hline
\multicolumn{1}{|c|}{{$g$}}&\multicolumn{1}{c}{ $r_c$ } & \multicolumn{1}{c}{ $P_c$}&\multicolumn{1}{c}{{$T_c$}}&\multicolumn{1}{c|}{$\epsilon$}  &\multicolumn{1}{c}{ $r_c$ } & \multicolumn{1}{c}{ $P_c$}&\multicolumn{1}{c}{{$T_c$}}&\multicolumn{1}{c|}{$\epsilon$} \\
\hline

\,\,\,\, $0.1$\,& \,\,\,\,\,\,\,1.15264\,\, &\,\,  0.01242\,\, &~~~~~~0.08024\,\,\,\,\,\,\,\,\,\,\,&\,\,0.28781& \,\,\,\,\,\,\,1.61595\,\, &\,\,  0.00628\,\, &~~~~~~0.05700\,\,\,\,\,\,\,\,\,\,\,&\,\,0.28708
\\
\
\,\, $0.2$\,\, & \,\,\,\,\,\,\,1.24344\,\, &\,\,  0.01124\,\,&0.07691\,\,\,\,\,\,\,\,\,\,\,&\,\,0.29350& \,\,\,\,\,\,\,1.66751\,\, &\,\,  0.00603\,\, &0.05606\,\,\,\,\,\,\,\,\,\,\,&\,\,0.28964
\\
\
\,\, $0.3$\,\, & \,\,\,\,\,\,\,1.41323\,\, &\,\,  0.00931\,\,&0.07080\,\,\,\,\,\,\,\,\,\,\,&\,\,0.30078& \,\,\,\,\,\,\,1.782557\,\, &\,\,  0.00551\, &0.05394\,\,\,\,\,\,\,\,\,\,\,&\,\,0.29440
\\
\
\,\, $0.4$\,\, & \,\,\,\,\,\,\,1.63033\,\, &\,\,  0.00739\,\,&0.06373\,\,\,\,\,\,\,\,\,\,\,&\,\,0.30646& \,\,\,\,\,\,\,1.95151\,\, &\,\,  0.00483\,\, &0.05089\,\,\,\,\,\,\,\,\,\,\,&\,\,0.29961      
\\
\
\,\, $0.5$\,\, & \,\,\,\,\,\,\,1.87393\,\, &\,\,  0.00581\,\,&0.05695\,\,\,\,\,\,\,\,\,\,\,&\,\,0.31032& \,\,\,\,\,\,\,2.15705\,\, &\,\,  0.00425\,\, &0.04740\,\,\,\,\,\,\,\,\,\,\,&\,\,0.30408
\\
 \hline
\end{tabular}
\end{center}
\caption{The critical horizon radius $r_c$, critical pressure $P_c$, critical temperature $T_c$ and universal constant $\epsilon$ for nonsingular--AdS EGB black holes.}
\label{tab3}
\end{table}
\end{center}
\end{widetext}
 \section{Critical Exponents}
 \label{critic}
 This section is devoted to the analysis of the behaviour of various thermodynamical quantities near the critical point that can be done by calculating the values of critical exponents  $\alpha, \beta, \gamma$, and $\delta$ \cite{Gunasekaran:2012dq}. We define following dimensionless quantities \cite{Kumar:2020cve,Nam:2019clw,Hyun:2019gfz}
\begin{eqnarray}\label{dimqun}
p=\frac{P}{P_c},~1+\tilde{\epsilon}=\frac{r_+}{r_c},~1+\omega=\frac{V}{V_c},~1+t=\frac{T}{T_c},
\end{eqnarray} to analyse the physical quantities near the critical point. We substitute the relation of $r_c$ from Eq. (\ref{dimqun}) in expression of black hole volume (\ref{V}) and get the relation between $\omega$ and $\tilde{\epsilon}$ 
\begin{equation}\label{46}
\omega=\frac{3}{r_c^3+g^3}\tilde{\epsilon}.
\end{equation}
 Rewriting the equation of state in terms of dimensionless quantities, we get
\begin{equation}\label{eqnst}
p=1+At-Bt\tilde{\epsilon}-C\tilde{\epsilon}^3+\mathcal{O}(t\; \tilde{\epsilon}^2,\; \tilde{\epsilon}^4),
\end{equation}
\begin{widetext}
with $A=T_c\left[\left(r_c^2+2\alpha \right)\left(r_c^3+g^3\right)\right]/(2P_c r_c^6)$, $B=T_c\left[\left(r_c^2+3\alpha\right)\left(r_c^3+2g^3\right)\right]/(P_c r_c^6)$ and\\ $C =\left[2r_c^3\left(-r_c^2+5\alpha\right)+g^3\left(3r_c^3+168\alpha\right)\\+2\pi  r_c T_c\left[r_c^3\left(r_c^2+20\alpha\right)+4g^3\left(5r_c^2+28\alpha\right)\right]\right]/(4 \pi P_c r_c^7)$.
\end{widetext}
 The critical exponents can be calculated by using following relations \cite{Kumar:2020cve,Tzikas:2018cvs}
 \begin{eqnarray}
C_V &=& T \frac{\partial S}{\partial T} \Big |_V \propto |t|^{-\alpha} \,,\;\;\;\;\; \label{crit_b}
\eta = V_l - V_s \propto |t|^{\beta} \,, \\ \label{crit_c}
\kappa_T &=& - \frac{1}{V} \frac{\partial V}{\partial P} \Big |_T \propto |t|^{-\gamma} \,,\; \label{crit_d}
|P-P_c|_{T=T_c}   \propto   |V-V_c|^{\delta}\nonumber \\,
\end{eqnarray}
where $V_l$ and $V_s$ are volumes of the large and small black holes. $C_V$, $\eta$ and $\kappa_T $ are the heat capacity at constant volume, the order parameter and the isothermal compressibility, respectively. The heat capacity at constant volume $C_V$ is zero for our system, and hence the value of $\alpha$ must be zero.
 To calculate the value of $\beta$, we use the well known Maxwell's area law \cite{Nam:2019clw,Hyun:2019gfz}
\begin{equation}\label{50}
\int_{\omega_s}^{\omega_l}\omega d\tilde{P}=0.
\end{equation}
By differentiating Eq. (\ref{eqnst}), we get
 \begin{equation}\label{51}
 dP=-P_c(Bt+3C\tilde{\epsilon}^2)d\tilde{\epsilon}.
 \end{equation}
 We use Eqs. (\ref{51}) and (\ref{46}) in Eq. (\ref{50}) and integrate to obtain
 \begin{equation}\label{52}
 \frac{1}{2}Bt\tilde{\epsilon}_l^2+\frac{3}{4}C \tilde{\epsilon}_l^4=\frac{1}{2}Bt\tilde{\epsilon}_s^2+\frac{3}{4}C \tilde{\epsilon}_s^4,
\end{equation}where $\tilde{\epsilon}_l$ and $\tilde{\epsilon}_s$, respectively, correspond the radii of large and small black hole. Because the pressure is not changing during the first-order phase transition of the small black holes to the large black holes, one can write 
\begin{equation}\label{49}
1+At-Bt\tilde{\epsilon}_{s} -C\tilde{\epsilon}_{s}^3=1+At-Bt\tilde{\epsilon}_{l} -C\tilde{\epsilon}_{l}^3.
\end{equation}
By solving Eqs. (\ref{52}) and (\ref{49}) simultaneously we get the following unique solution
\begin{equation}
\tilde{\epsilon}_l=-\tilde{\epsilon}_s=\sqrt{\frac{-Bt}{C}}.
\end{equation}
Thus Eq. (\ref{crit_b}), takes the form
\begin{equation}
\eta=V_l-V_s=V_c(\omega_l-\omega_s)\propto \sqrt{-t},
\end{equation} which gives $\beta=1/2$. Further, we calculate the thermal compressibility 
\begin{equation}
\kappa_T = - \frac{1}{V} \frac{\partial V}{\partial P}|_{T}=\frac{3}{P_c(1+\omega)(r_c^3+g^3)}(\frac{\partial \tilde{\epsilon}}{\partial p}|_t)\propto\frac{1}{Bt},
\end{equation} 
to get $\gamma=1$. The critical isotherms at $t=0$, become 
\begin{equation}
P=P_c (1-C\tilde{\epsilon}^3)~~~~~~~\textit{or}~~~~~~~|P-P_c|_{T_c}\varpropto \tilde{\epsilon}^3\propto|V-V_c|^3,
\end{equation} which leads us to get $\delta=3$. The above calculations of critical exponents for nonsingular--AdS black holes in $4D$ EGB gravity confirm that these are the same as those of Van der Walls fluid. 
\section{Conclusion}
\label{con}
Black hole thermodynamics has received significant attention in the last decade, as the researchers have explored the understanding of the cosmological constant $\Lambda$ as the thermodynamic pressure. A critical conceptual development is understood that the mass term $M$ in the First Law of thermodynamics, is treated as enthalpy of the system. It revealed several new interesting properties for semiclassical black holes analogous to known chemical phenomena, such as $P–V$ criticality for charged AdS black holes.  Despite the thermodynamic correspondence with Van der Waals fluids are well established, many problems and open questions remain to be explored, e.g., it would be interesting to see these phenomena in more complicated higher curvature EGB gravity coupled NED. Motivated by this, we analysed the thermodynamics and phase transition of the nonsingular black holes inside an extended AdS phase space in $4D$ EGB gravity with higher curvature terms.  From a geometrical point of view, we have derive exact static spherically symmetric Hayward-like  AdS  black holes in regularized $4D$ EGB gravity, i.e., the nonsingular-AdS EGB black holes (\ref{fr}) which exactly encompassed $4D$ EGB AdS black holes in the absence of NED ($g \to 0$) and the Schwarzschild AdS black holes when  $\alpha,\; g \to 0$.  The nonsingular $4D$ EGB AdS black holes (\ref{fr}), subject to constraints on the parameters, admit horizons which could be at most two, describing a variety of charged, self-gravitating objects, including an extremal black hole with degenerate horizons and a nonextremal black hole with Cauchy and event horizons.  

We have presented the global thermodynamic properties, including thermodynamic stability and phase transition of nonsingular-AdS EGB black holes in extended phase space, to analyze the effect of higher curvature terms and NED.  From the higher curvature terms and NED corrected temperature, it turns out that the black hole no longer evaporates completely but ends at zero temperature. The entropy is corrected by an additional logarithmic term which reflects the effect of higher curvature gravity and also correction term due to NED. Through the calculation and plots of the heat capacity, it is shown that there is a critical point where the heat capacity diverge and marking the phase transition. The divergence of the thermodynamic curvature means a phase transition indeed happens when the Hawking temperature takes extrema. 

Our analysis revealed the analogy between the nonsingular--AdS EGB black holes and Van der Waals liquid-gas system. The heat capacity at constant pressure showed discontinuities and hence indicated the existence of second-order phase transition for pressure less than critical value. From the free energy, we gave a global analysis of the stability of the black hole, which lead to find that at pressure less than critical pressure, the black hole can undergo first-order phase transition to large black hole.We showed that the larger black hole is more stable than the small one, which means a Hawking–Page-like phase transition. While studying the isotherms on $P-V$ planes, we found that the black hole exhibited Van der Walls-like phase transition for black hole temperature less than critical value. Last but not least, we calculated the values of critical exponents to see the behaviour of thermodynamical quantities near the critical point to found out that our system behaves like Van der Walls fluid.

\subsection*{Acknowledgments}
S.G.G. would like to thank SERB-DST for the ASEAN project IMRC/AISTDF/CRD/2018/000042. S.G.G. would like to also thank IUCAA, Pune for the hospitality while this work was being done.

\end{document}